\documentclass[aps, prl, twocolumn]{revtex4}
\usepackage{graphicx}
\usepackage{dcolumn}
\usepackage{bm}
\usepackage{natbib}

\begin{document}

\title{Comment to "Mechanism for Designing Metamaterials with a High Index of Refraction"}

\author{A. M. Shuvaev}
\author{A. Pimenov}
\affiliation{Experimentelle Physik IV, Universit\"{a}t W\"{u}rzburg,
97074 W\"{u}rzburg, Germany} %

\date{\today}


\maketitle

\begin{figure}[t]
\includegraphics[angle=0,width=0.73\linewidth,clip]{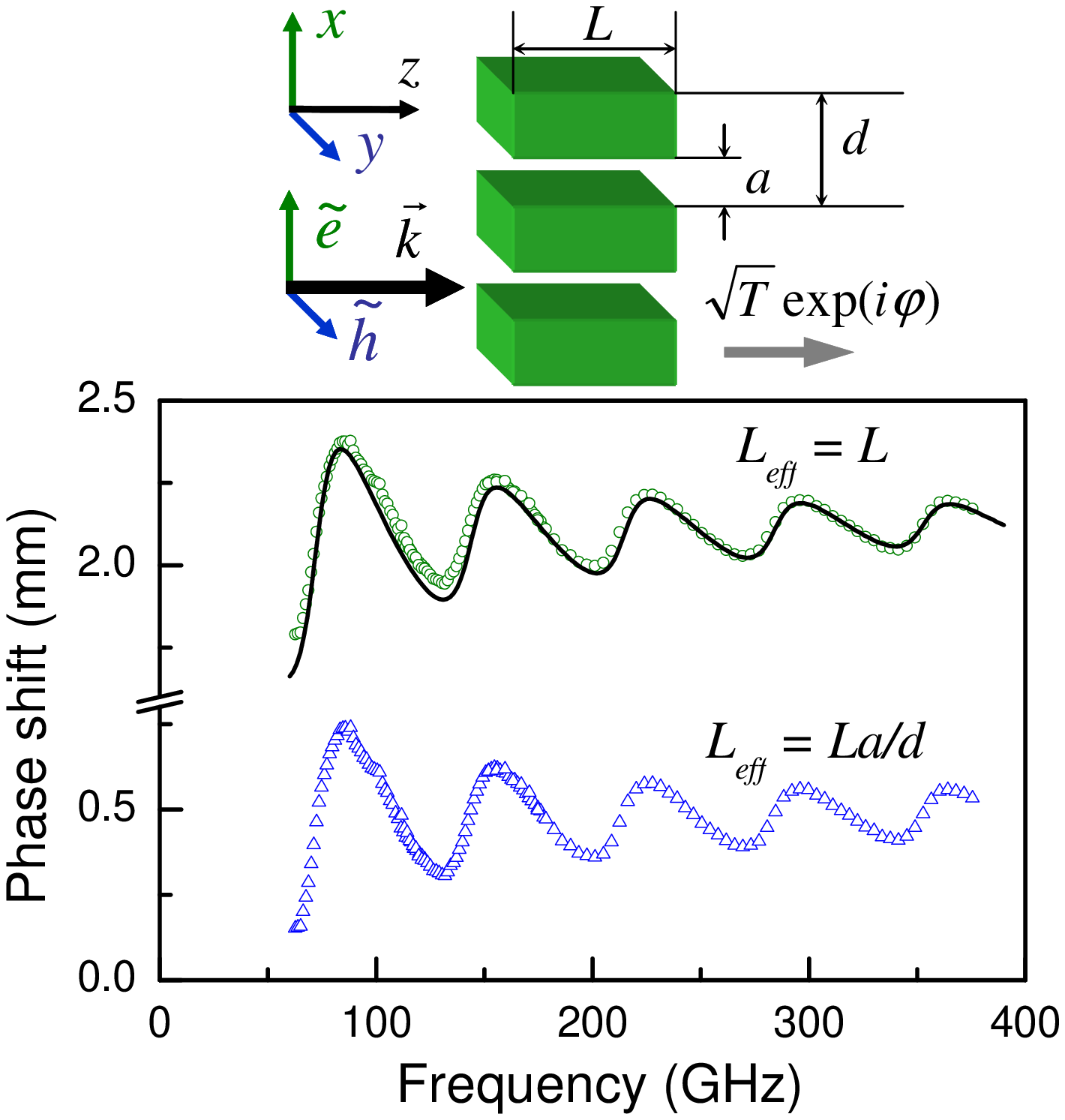}
\caption{Experimental phase shift $\phi \lambda / 2 \pi$ of the
metamaterial shown in the inset. Symbols represent the experimental
data from Ref.~\protect\cite{wir}. Green circles - sample thickness is taken
as $L$ (this Comment); blue triangles - sample thickness is taken as
$La/d$ (Ref.~\protect\cite{shen}). Both models give the same theoretical
values for the phase shift (black solid line), but influences the
experiment in different ways. The parameters are: $L=2.01$ mm,
$d/a=6$ (the fits utilizes $(d/a)_{eff}=5.5$). Inset shows the
geometry of the sample and experiment. The sample is supposed to be
infinite in the $y$-direction.}
\label{fig1}
\end{figure}

In a recent Letter~\cite{shen} Shen et al. presented a mechanism for
designing metamaterials with a high index of refraction. The
proposed metamaterial consists of a metal with tiny slits in it
(Fig.~\ref{fig1}). This system is suggested to be equivalent to a
dielectric with an effective refractive index of $n_{eff}= d/a$ and
effective thickness $L_{eff}= L/n_{eff}$.

In an attempt to prove these ideas experimentally, one of us (AP)
carried out transmittance and reflectance experiment of the metallic
metamaterial with slits~\cite{wir}. In these experiments the
predictions of the theory of Ref.~\cite{shen} concerning the
amplitudes of the transmittance and reflectance have been confirmed.
However, the discrepancy of the transmittance phase shift with the
predictions of the model has been observed: the experimental phase
had to be corrected by a value $\phi_{corr}=2\pi (L-L_{eff})/\lambda$,
with $\lambda$ being the wavelength of the radiation.

In this Comment we suggest another explanation for the observed
effects. Namely, the metamaterial possesses an effective dielectric
permittivity equal to $\varepsilon_{eff} = d/a$ and an effective
magnetic permeability $\mu_{eff} = a/d$. Therefore, this
metamaterial is in fact a low-impedance material with $z_{eff}/z_0 = a/d$
and the refractive index equal to unity, $n_{eff}= 1$. Here
$z_0= \sqrt{\mu_0/ \varepsilon_0}$ is the wave impedance of the free
space.

In order to demonstrate this idea we consider a quasi-static
approximation with the electric fields ($E$) perpendicular to
the slits and magnetic fields ($H$) parallel to the slits
(Fig.~\ref{fig1}). In the following we have dropped the corresponding
indices and use the field averaging over the unit
cell~\cite{pendry}. For electric fields within the metamaterial we
utilize the boundary condition in which the normal component of the
electric displacement $D=\varepsilon_0 \varepsilon E$ is continuous
at the slit boundary. Taking into account that: i) $D$ remains
constant within the unit cell, ii) in the air slit $D=\varepsilon_0 E$,
and iii) within the metal $E=0$, we get for the average electric
field $\varepsilon_0 \langle E \rangle / D =\varepsilon_{eff}^{-1}= a/d$.
The effective magnetic permeability of the metamaterial may be
calculated in a similar way using the continuity of the tangential
component of the magnetic field $H$. In that case the flux density
$B=\mu_0\mu H$ and not the field $H$ is zero within the metal. We
get $\langle B\rangle/\mu_0 H=\mu_{eff}= a/d$. Similar expressions
for permittivity and permeability have been obtained in Ref.~\cite{garcia}.

The ability of the present model to describe the experimental data
is demonstrated in Fig.~\ref{fig1}, which compares the phase shift
$\phi$ as obtained within both models. The most important point here
is that the sample thickness should be known during the measurements
and, therefore, the \emph{model assumptions influence already the
experimental values}. Because the sample thickness is different in
both models, the data vary by $\Delta(\phi \lambda / 2 \pi) = L - L_{eff}$.

Finally we note that the amplitudes of the reflectance and
transmittance calculated in Ref.~\cite{shen} remain correct, because
the relevant reflection coefficient at the metamaterial surface is
equal to $r=(1-z)/(1+z) = (d-a)/(d+a)$,  i.e. the same expression as
used in Ref.~\cite{shen}. The positions of the Fabry-P\'{e}rot
resonances remains the same as well, because the optical thickness
of the metamaterial $L_{opt}=nL_{eff}$ is the same in both models.

\end{document}